\documentclass{research4cacm}

\usepackage{enumitem}
\usepackage[utf8]{inputenc}
\usepackage{epsfig,endnotes}
\usepackage{varwidth}
\usepackage{algorithm}
\usepackage{algpseudocode}
\usepackage{balance}
\usepackage{nicefrac}
\usepackage{amsmath}
\usepackage{mathtools}
\usepackage{multirow}
\usepackage{bigdelim}
\usepackage{mathtools}
\usepackage{amssymb}
\usepackage{indentfirst}
\usepackage{booktabs}
\usepackage{enumitem}
\usepackage{float}
\usepackage{graphicx}
\usepackage{caption, subcaption}
\usepackage[normalem]{ulem}
\usepackage[titletoc]{appendix}
\usepackage{flushend}
\usepackage{subcaption}
\usepackage[normalem]{ulem}
\usepackage{xpatch}
\usepackage{stmaryrd}
\usepackage{MnSymbol}
\usepackage{xspace}
\usepackage{amssymb}
\usepackage{framed,mdframed}
\usepackage[dvipsnames]{xcolor}
\usepackage[breaklinks]{hyperref}
\usepackage{csquotes}

\newcommand{\squishlist}{
	\begin{list}{$\bullet$}
		{
			\setlength{\itemsep}{0pt}
			\setlength{\parsep}{3pt}
			\setlength{\topsep}{3pt}
			\setlength{\partopsep}{0pt}
			\setlength{\leftmargin}{1.5em}
			\setlength{\labelwidth}{1em}
			\setlength{\labelsep}{0.5em} } }
	
	\newcommand{\squishend}{
\end{list}  }

\newcommand{\Term}{DP-cryptography\xspace}
\newcommand{\termic}{DP-cryptographic\xspace}
\newcommand{\TERM}{DP-Cryptography\xspace}

\newcommand{\eat}[1]{}

\graphicspath{{./Images/}}
\DeclareGraphicsExtensions{.pdf,.jpg,.png}

\newtheorem{theorem}{Theorem}

\newtheorem{definition}[theorem]{Definition}

\newcommand{\namedref}[2]{\hyperref[#2]{#1~\ref*{#2}}\xspace}

\newcommand{\etal}{\textit{et. al. }}

\newcommand{\ch}[1]{#1}
\newcommand{\chh}[1]{#1}

\newcommand{\comment}[1]{}

\newcommand{\xh}[1]{#1}

\begin{document}

% --- Author Metadata here ---
% Conference information is NOT appropriate for CACM so comment it out.
\CopyrightYear{2020} % Allows default copyright year (2008) to be over-ridden - IF NEED BE.
%\crdata{0001-0782/08/0X00}  % Allows default copyright data (0001-0782/08/0X00) to be over-ridden - IF NEED BE.	
% --- End of Author Metadata ---

\title{
%From Differential Privacy to Leaky Cryptography: Getting Accurate and Fast without Being Perfect
\TERM: Marrying Differential Privacy and Cryptography in Emerging Applications
%From Differential Privacy to Leaky Cryptography: Emerging Directions in Privacy-Preserving Data Analytics
%Emergence of Differentially Private Cryptographic Primitives\\ New Directions in Differentially Private Cryptography
% \titlenote{(This is a simple titlenote.)For use with research4cacm.cls. Supported by ACM.}
%
% Show use of \thanks - which can appear here (normal/default) or down by the author
%\thanks{The original version of this paper is entitled ``XXX" and was
%published in (Title of publication, publication date, publisher.)}
}
%\subtitle{[Extended Abstract]
%\titlenote{A full version of this paper is available in...}
% and affiliations.
%
\numberofauthors{4} %  in this sample file, there are a *total*
% of SIX authors and all of them fit neatly on the first page.
% As said, all authors get 'equal billing' and you should fit all of them on the opening page
% in the 'byline'. The production/editorial-staff will 'separate' names from their affiliations, leaving
% author names beneath the title (in the byline), and moving the affilations/contact information to an area
% after the references at the back of the article.
%
\author{
% 1st. author
\alignauthor Sameer Wagh\\
       \affaddr{Princeton University}\\
       \email{swagh@princeton.edu}
% 2nd. author
\alignauthor Xi He\\
       \affaddr{University of Waterloo}\\
       \email{xihe@uwaterloo.ca}
% 3rd. author
\and 
\alignauthor Ashwin Machanavajjhala\\
       \affaddr{Duke University}\\
       \email{ashwin@cs.duke.edu}
%\and  % use '\and' if you need 'another row' of author names
% 4th. author
\alignauthor Prateek Mittal\\
       \affaddr{Princeton University}\\
       \email{pmittal@princeton.edu}
}

\maketitle
\section*{Abstract}
%Differential privacy (DP) has arisen as the state-of-the-art metric for quantifying individual privacy when sensitive data are analyzed, and it is increasingly being deployed in statistical organizations (like the US Census Bureau) and the industry (Apple, Google and Microsoft). 
\ch{Differential privacy (DP) has arisen as the state-of-the-art metric for quantifying individual privacy when sensitive data are analyzed, and it is starting to see practical deployment in organizations such as the US Census Bureau, Apple, Google, Facebook, and Microsoft.}
There are two popular models for deploying differential privacy -- standard differential privacy (SDP), where a trusted server aggregates all the data and runs the DP mechanisms, and local differential privacy (LDP), where each user perturbs their own data and perturbed data \ch{is} analyzed. Due to security concerns arising from aggregating raw data at a single server, several real world deployments in industry have embraced the LDP model~\cite{rappor, appleDP,facebookDP, uberDP}. However, systems based on the LDP model tend to have poor utility -- ``a gap'' in the utility achieved as compared to systems based on the SDP model.

In this work, we survey and synthesize emerging directions of research at the intersection of differential privacy and  cryptography. First, we survey solutions that combine cryptographic primitives like secure computation, anonymous communication and oblivious computation with differential privacy to give alternatives to the LDP model that avoid a trusted server as in SDP but close the gap in accuracy. These cryptographic primitives introduce performance bottlenecks and necessitate efficient alternatives. Second, we synthesize work in an area that we call ``DP-Cryptography" -- cryptographic primitives that are allowed to leak differentially private outputs. These primitives have orders of magnitude better performance than standard cryptographic primitives. \Term primitives are perfectly suited for implementing alternatives to LDP, but are also applicable to scenarios where standard cryptographic primitives do not have practical implementations. Through this unique lens of research taxonomy, we survey the landscape of ongoing research in these directions while also providing novel directions for future research.

\section{Introduction}\label{sec:introduction}
On Feb 15, 2019, John Abowd, chief scientist at the US Census Bureau, announced the results of a \textit{reconstruction attack} that they proactively launched using data released under the 2010 Decennial Census~\cite{garfinkel2018understanding}. The decennial census released billions of statistics about people like “how many people of the age 10-20 live in New York City” or “how many people live in 4 person households”. Using only the data publicly released in 2010, an internal team was able to (a) correctly reconstruct records of address (by census block), age, gender, race and ethnicity for 142 million people (about 46\% of the US population), and (b) correctly match these data to commercial datasets circa 2010 to associate personal-identifying information such as names for 52 million persons (17\% of the population). This is not specific to the US Census Bureau -- such attacks can occur in any setting where statistical information in the form of deidentified data, statistics or even machine learning models are released. That such attacks are possible was predicted over 15 years ago by a seminal paper by Irit Dinur and Kobbi Nissim \cite{dinur2003revealing} -- releasing a sufficiently large number of aggregate statistics with sufficiently high accuracy provides sufficient information to reconstruct the underlying database with high accuracy. The practicality of such a large scale reconstruction by the US Census Bureau underscores the grand challenge that public organizations, industry, and scientific research faces: how can we safely disseminate results of data analysis on sensitive databases?  

An emerging answer is \textit{differential privacy}. An algorithm satisfies differential privacy (DP) if its output is insensitive to adding, removing or changing one record in its input database. Differential privacy is considered the ``gold standard" for privacy for a number of reasons. It provides a persuasive mathematical proof of privacy to individuals with several rigorous interpretations \cite{pufferfish,kasiviswanathan2008thesemantics}. The differential privacy guarantee is composable and repeating invocations of differential private algorithms lead to a graceful degradation of privacy. \ch{The US Census Bureau was the first big organization to adopt differential privacy in 2008 for a product called OnTheMap~\cite{ashwin08:icde}, and subsequently there have been deployments by Google, Apple, Microsoft, Facebook, and Uber~\cite{rappor, appleDP, telemetry17, facebookDP, uberDP}.}%\footnote{\ch{Though these deployments have had limitations, these systems are constantly being improved for better deployments.}}.} %intend to use differentially private algorithms for the 2020 Decennial release. 

Differential privacy is typically implemented by collecting data from individuals \textit{in the clear} at a \textit{trusted data collector}, then applying one or more differentially private algorithms, and finally releasing the outputs. This approach, which we call \textit{standard differential privacy (SDP)}, works in cases like the US Census Bureau where there is a natural trusted data curator. However, when Google wanted to monitor and analyze the Chrome browser properties of its user base to detect security vulnerabilities, they chose a different model called \textit{local differential privacy (LDP)}. In LDP, individuals perturb their records \textit{before} sending it to the server, obviating the need for a trusted data curator. Since the server only sees perturbed records, there is no centralized database of sensitive information that is susceptible to an attack or subpoena requests from governments. The data that Google was collecting -- browser fingerprints -- uniquely identify individuals. By using LDP, Google was not liable to storing these highly identifying user properties. Due to these attractive security properties a number of real world applications of differential privacy in the industry -- Google's RAPPOR~\cite{rappor},  Apple Diagnostics~\cite{appleDP} and Microsoft Telemetry~\cite{telemetry17} -- embrace the LDP model. 

However, the improved security properties of LDP come at a cost in terms of utility. Differentially private algorithms hide the presence or absence of an individual by adding noise. Under the SDP model, counts over the sensitive data, e.g., ``number of individuals who use the \url{bing.com} search engine", can be released by adding a constant amount of noise. In the LDP model, noise is added to \textit{each individual record.} Thus, answering the same count query requires adding $O(\sqrt{N})$ error \ch{(Theorem 2.1 from~\cite{Cheu2018DistributedDP})} for the same level of privacy, where $N$ is the number of individuals participating in the statistic. In other words, under the LDP model, for a database of a billion people, one can only learn properties that are common to at least 30000 people \ch{($O(\sqrt{N})$)}. In contrast, under SDP, one can learn properties that are shared by as few as a 100 people \ch{($O(1)$ including constants; cf~\cite{DP})}. Thus, the LDP model operates under more practical trust assumptions than SDP, but as a result incurs a significant loss in data utility. In this work, we review literature in this domain under two categories:

%To bridge this gap between SDP and LDP, there has been a lot of recent interest in the research community to achieve the best of both worlds -- operate in a model with fewer trust assumptions yet achieve the high utility comparable to a deployment in the SDP model. In this paper, we review this literature and synthesize them under two categories: 

\squishlist
\item \textbf{Cryptography for DP:} We review a growing line of research that aims to use cryptographic primitives to bridge the gap between SDP and LDP. In these solutions, the trusted data curator in SDP is replaced by cryptographic primitives that result in (a) more practical trust assumptions than the SDP model, and (b) better utility than under the LDP model. Cryptographic primitives such as anonymous communication and secure computation have shown significant promise in improving the utility of differentially private implementations while continuing to operate under the practical trust assumptions that are accepted by the security community.
\item \textbf{DP for Cryptography:} Differential privacy is typically applied to settings that involve complex analytics over large datasets. Introducing cryptographic primitives results in concerns about the feasibility of practical implementations at that scale. This has given rise to a second line of work that employs differential privacy as a tool to speed up cryptographic primitives, thereby pushing the frontiers of their practical deployments. While the original cryptographic primitives are defined with respect to perfect privacy, under differential privacy, it is ok to learn distributional information about the underlying dataset. We explore in depth the following cryptographic primitives (a) secure computation %(2) secure data access 
(b) secure communication, and show how in the context of differential privacy one can build ``leaky" but efficient implementations of these primitives.
\squishend

\ch{These lines of work both reflect exciting directions for the computer science community. We begin by giving a brief technical introduction to differential privacy in Section~\ref{sec:dp}. 
%Next, in Section~\ref{sec:dppractice}, we examine two popular deployment scenarios for differential privacy -- SDP and LDP. 
We discuss the ``Cryptography for DP'' paradigm in Section~\ref{sec:cryptofordp} and ``DP for cryptography'' in Section~\ref{sec:dpforcrypto}. Section~\ref{sec:open} provides concrete ideas for future work as well as open problems in the field through the lens of combining differential privacy and cryptography. }\\

\hspace*{-\parindent}%
\fbox{\begin{minipage}{0.95\columnwidth}
\textbf{Key Insights}\vspace{1mm} \\
$\bullet$ Local Differential Privacy is increasingly being embraced as the primary model of deployment of differential privacy, albeit at a heavy accuracy cost.\\ 
$\bullet$ Cryptographic primitives can help bridge the utility gap between systems deployed in the local differential privacy model and standard differential privacy model but the increased utility may come at the cost of performance. \\
$\bullet$ \termic primitives, that are relaxed notions of cryptographic primitives that leak differentially private outputs, permit implementations that are orders of magnitude faster than the regular primitives.
\end{minipage}}

\section{Differential Privacy}\label{sec:dp}

Differential privacy~\cite{differentialprivacy} is a state-of-the-art privacy metric for answering queries from statistical databases while protecting individual privacy. Since its inception, there has been considerable research in both the theoretical foundations~\cite{dinur2003revealing, dwork2006calibrating} as well as \ch{some} real world deployments~\cite{rappor, appleDP} of differential privacy. The rigorous mathematical foundation and the useful properties of differential privacy have led to an emerging consensus about its use among the security and privacy community.

\subsection{Definition of Differential Privacy}
Informally, the privacy guarantees of differential privacy can be understood as follows: Given any two databases, otherwise identical except one of them contains random data in place of data corresponding to any \textit{single} user, \chh{differential privacy requires} that the response mechanism will behave approximately the same on the two databases. Formally,

\begin{definition}
Let $M$ be a randomized mechanism that takes a database instance $D$ and has a range $\mathcal{O}$. We say $M$ is $(\epsilon,\delta)$-differentially private, if for any neighboring databases $(D_1,D_2)$ that differ in the data of a single user, and for any $S\subseteq O$, we have 
\begin{equation} \label{eq:epsilondeltaDP}
\Pr[M(D_1) \in S] \leq e^\epsilon \Pr[M(D_2) \in S] + \delta
\end{equation}
\end{definition}

%This $(\epsilon, \delta)$-differential privacy definition characterizes the effect of a small change in the input database on the change in the distribution of the output (hence the term differential). In other words, if the input is perturbed slightly, then with high probability, the output probability distribution changes only slightly. %The $\delta$-term in Eq.~\ref{eq:epsilondeltaDP} is a measure of the failure probability, and when $\delta = 0$ the definition reduces to pure $\epsilon$-differential privacy. 

\ch{Differential privacy enjoys some important properties that make it a useful privacy metric.} First, the privacy guarantees of differential privacy have been thoroughly studied using various metrics from statistics and information theory such as hypothesis testing and Bayesian inference~\cite{pufferfish,kasiviswanathan2008thesemantics}. %\am{see citations in intro}. 
Thus, the semantic meaning of its privacy guarantees is well understood.
%Below, we state three of the most important properties of differential privacy:
%
%\begin{theorem}{\textbf{Safety against Post Processing:}}\label{thm:postprocessing}
%Let $M$ be an $(\epsilon, \delta)$-differentially private mechanism. Let $f$ be an arbitrary mapping from the output of the mechanism to any other domain. Then the composition $f \circ M$ is also $(\epsilon, \delta)$-differentially private.
%\end{theorem}
%
%\begin{theorem}{\textbf{Sequential Composition:}}\label{thm:sequential}
%Given two independent mechanisms $M_1, M_2$ with privacy guarantees $(\epsilon_1,\delta_1)$ and $(\epsilon_2,\delta_2)$, the composite mechanism obtained by sequentially applying $M_1$ and $M_2$ is a $(\epsilon_1+\epsilon_2,\delta_1+\delta_2)$-differentially private mechanism.
%\end{theorem}
%
%\begin{theorem}{\textbf{Parallel Composition:}}\label{thm:parallel}
%Given two independent mechanisms $M_1, M_2$ with privacy guarantees $(\epsilon_1,\delta_1)$ and $(\epsilon_2,\delta_2)$ which compute on disjoint subsets of a private database, the overall function is $\left( \max(\epsilon_1, \epsilon_2), \max(\delta_1, \delta_2) \right)$-differentially private.
%\end{theorem}
%Secondly, differential privacy has a number of composition properties which enable the analysis of privacy leakage in many compound scenarios such as multiple invocations, post-processing etc. %three important properties -- sequential composition, parallel composition, and safety against post-processing. These properties are very useful in proving the privacy guarantees of complex algorithms.
Differential privacy also has a number of composition properties which enable the analysis of privacy leakage for complex algorithms. In particular, sequential composition addresses the impossibility result by Dinur and Nissim~\cite{dinur2003revealing} and quantifies the degradation of privacy as the number of sequential accesses to the data increases. The post-processing theorem (a special case of sequential composition) ensures that the adversary cannot weaken the privacy guarantees of a mechanism by transforming the received response. The end-to-end privacy guarantee of an algorithm over the entire database can thus be established using the above composition theorems and more advanced theorems~\cite{DP}.

%Sequential composition allows the quantification of privacy loss after multiple invocations of a differentially private mechanism. On the other hand, parallel composition allows the quantification of privacy loss when a differentially private mechanism acts on disjoint subsets of a private database. Finally, the post-processing theorem ensures that the adversary cannot weaken the privacy guarantees of a mechanism by transforming the received response. The end-to-end privacy guarantee of an algorithm over the entire database can thus be established using the above theorems.

%Moreover, the sequential composition theorems address the impossibility result by Dinur and Nissim~\cite{dinur2003revealing} and quantifies the degradation of privacy as the number of sequential accesses to the data increases. There are more advanced composition theorems that enable a tighter analysis of the privacy loss under various scenarios such as when the queries are not adaptively chosen~\cite{differentialprivacy}. For more in depth reading, we refer the reader to the differential privacy paper by Dwork \textit{et al.}~\cite{DP}.

\begin{figure}
\includegraphics[width=\columnwidth]{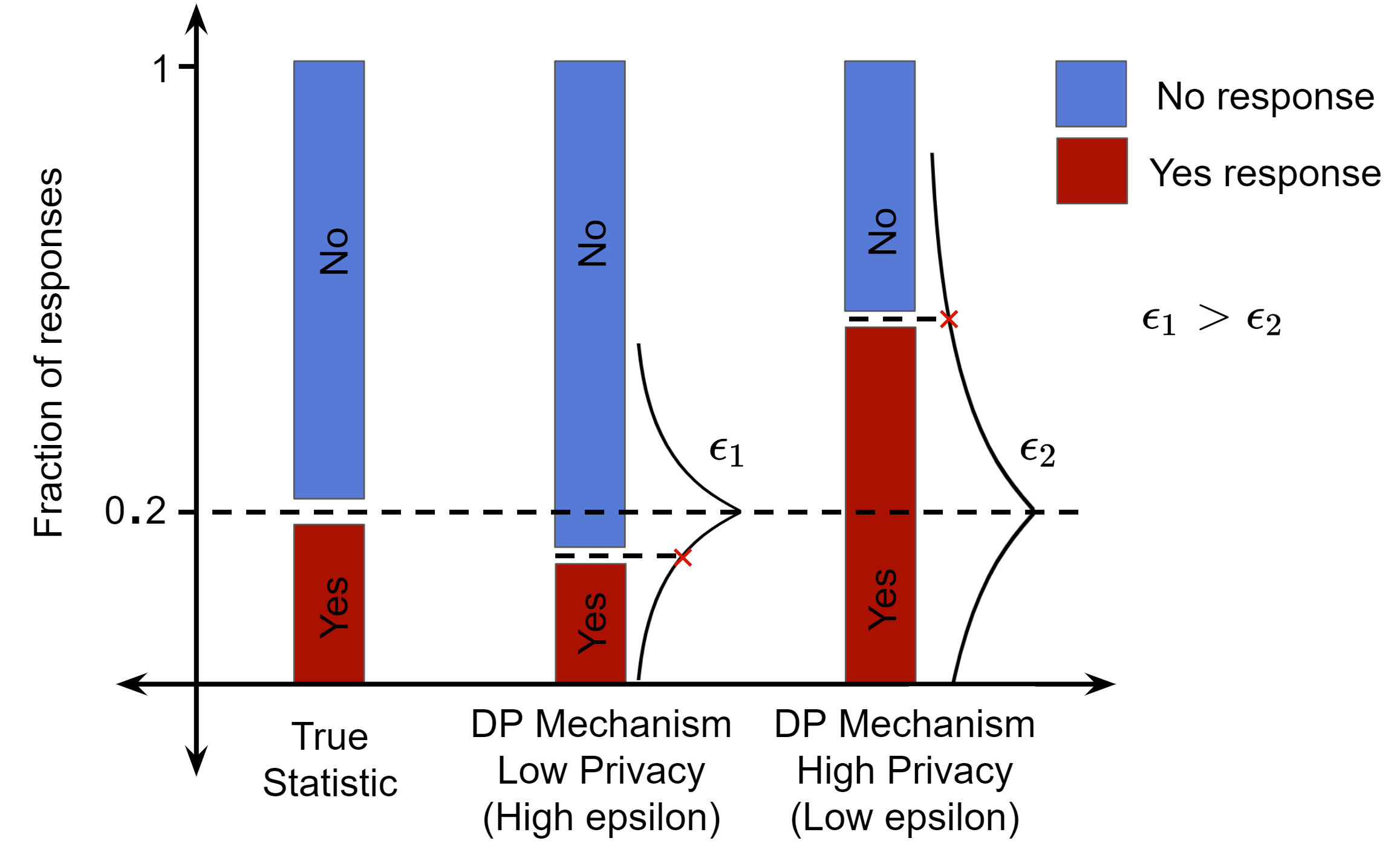}
\caption{Differentially private mechanisms randomize query response to achieve privacy. If the true response to a query such as ``What fraction of users use drugs illegally?'' \ch{was 20\%}, then a high privacy response mechanism (low $\epsilon$ value) will add a lot of noise yielding low utility. On the contrary, if a low privacy response mechanism was used (high $\epsilon$ value), the response will be very close to 20\% yielding high utility.}\label{fig:RRandLM}
\end{figure}

\subsection{Differentially Private Mechanisms}\label{subsec:mechanisms}
Next, we review two classic differentially private mechanisms, the Laplace mechanism and the Randomized Response mechanism, with the following scenario: a data analyst would like to find out how many users use drugs illegally. Such a question would not elicit any truthful answers from users and hence we require a mechanism that guarantees (a) response privacy for the users and (b) good utility extraction for the data analyst. 

\textbf{Laplace Mechanism: }The Laplace mechanism~\cite{differentialprivacy} considers a trusted data curator (SDP model) who owns a table $N$ of truthful records of users, for example, each record indicates whether a user uses drugs illegally. If a data analyst would like to learn how many users use drugs illegally, the data curator (trusted) computes the true answer of this query and then perturbs it with a random (Laplace distributed) noise that is sufficient to provide privacy. The magnitude of this noise depends on the largest possible change on the query output -- also known as the sensitivity of the query -- if the data corresponding to a single user is changed. 

\textbf{Randomized Response Mechanism: }Randomized response was first introduced by Warner in 1965 as a research technique for survey interviews. It enabled respondents to answer sensitive questions (about topics such as sexuality, drug consumption) while maintaining the confidentiality of their responses. An analyst interested in learning aggregate information about sensitive user \chh{behavior % such as the number of users using drugs illegally 
would} like to query this function on a database that is \textit{distributed} across 
$N$ clients with each client having its own private response $x_1, \hdots ,x_N$. Instead of releasing $x_i$ directly, the clients release a 
perturbed version of their response $y_i$, thus maintaining response privacy. The analyst collects these perturbed responses and recovers meaningful statistics using reconstruction techniques. %Perturbing the responses provides the respondents with a leeway to assert that they were not truthful in their response.

Both these approaches have gained popularity in many applications of differential privacy due to their simplicity as well as the rigorous privacy guarantee on user data. Fig.~\ref{fig:RRandLM} shows the behavior of differentially private mechanisms for two different privacy values in reference to the true statistic. A less private response results in a more accurate query result while a more private response results in a less accurate query result.

\ch{

%\section{Trust Assumptions in the Deployment of DP}
\section{Cryptography for Differential Privacy}
\label{sec:cryptofordp}

\begin{figure*}[!th]
\centering
\includegraphics[width=\linewidth]{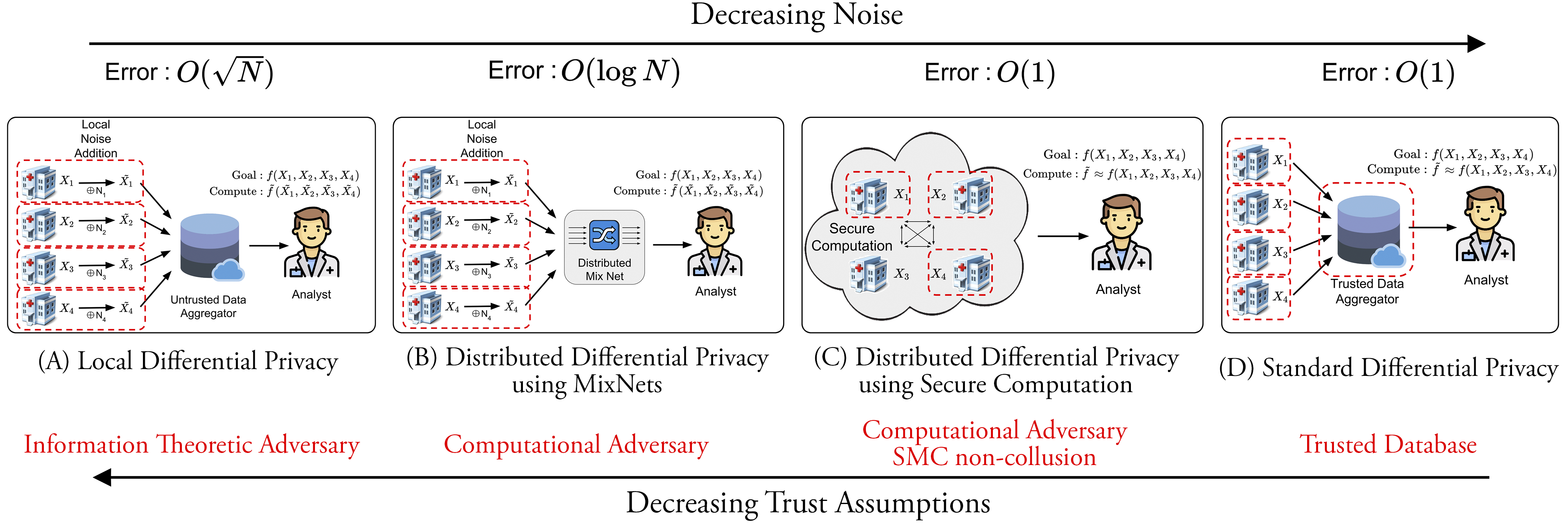}
\caption{This figure shows various deployment scenarios of differential privacy and the underlying trust assumptions in each of them. Standard Differential Privacy (SDP, Fig.~\ref{fig:crypto_for_dp}D) assumes a trusted database, and is thus able to achieve high accuracy i.e., $O(1)$ error. Local Differential Privacy (LDP, Fig.~\ref{fig:crypto_for_dp}A) on the other hand, does not rely on the use of a trusted database but achieves lower accuracy i.e., $O(\sqrt{N})$ error. The goal is to achieve utility of the SDP setting while operating under more practical assumptions such as the LDP setting (i.e., no trusted database). Fig~\ref{fig:crypto_for_dp}B and Fig~\ref{fig:crypto_for_dp}C show how different cryptographic primitives can be used to improve the utility of DP deployments under such practical assumptions.}
\label{fig:crypto_for_dp}
\end{figure*}

By itself, differential privacy is a guarantee on a mechanism and hence is ``independent'' of the deployment scenario. However, when used in practice, practical trust assumptions are made that enable the deployment of differential privacy based systems. In this section, we consider two popular 
deployment scenarios for differential privacy -- Standard Differential Privacy (SDP, graphically represented in Fig.~\ref{fig:crypto_for_dp}D) and Local Differential Privacy (LDP, graphically represented in Fig.~\ref{fig:crypto_for_dp}A). SDP relies on the need for a trusted data aggregator who follows the protocol. However, in practice, a trusted data aggregator may not always exist. LDP on the other hand does not require a trusted data aggregator\footnote{\ch{Differentially private federated learning is simply a special case of the LDP deployment scenario.}}. With privacy regulations such as GDPR and FERPA, large organizations such as Google increasingly embrace the LDP model thereby avoiding the liability of storing such sensitive user data. This approach also insures data collectors from potential theft or subpoenas from the government. For these reasons, LDP is frequently a more attractive deployment scenario. However, the utility of the statistics released in LDP is poorer than that in SDP. Consequently, there is a gap in the trust assumptions and the utility achieved by mechanisms in SDP and LDP: high trust assumptions, high utility in SDP and lower trust assumptions, lower utility in LDP. We ask the following question: 
%Because of the lower trust assumption in LDP, the utility of the statistics released is poorer than that in SDP. 
%Avoiding centralized collection of truthful data also insures data collectors from potential theft or subpoenas from the government.  

\begin{displayquote}
Can cryptographic primitives help in bridging the gap that exists between mechanisms in the standard differential privacy model and the local differential privacy model? 
\end{displayquote}
An emerging direction of research has been to explore the use of cryptography to bridge the trust-accuracy gap and obtain the best of both worlds: high accuracy without assuming trusted data aggregator. In this section, we explore in depth two concrete examples of the role of cryptography in bridging this gap (a) anonymous communication (b) secure computation and encryption. %to obtain the best of these deployment scenarios.

\textbf{Key Challenges: }There exists a big gap in the accuracy and trust achieved by known mechanisms in the standard differential privacy setting with a trusted data curator (Fig.~\ref{fig:crypto_for_dp}D) and local differential privacy without such a trusted curator (Fig.~\ref{fig:crypto_for_dp}A). Achieving the utility as in the SDP setting while operating under practical trust assumptions such as those in LDP has proven to be a tough challenge. Cryptographic primitives show promise in solving this challenge.\\ %bridging this utility gap.

\vspace{2mm}
\hspace*{-\parindent}%
\fbox{\begin{minipage}{0.95\columnwidth}
		\textbf{Key Insights of using Cryptography for DP}\vspace{1mm} \\
$\bullet$ Increasing privacy regulations such as GDPR and FERPA have pushed organizations such as Google to embrace the LDP model for deployment of differential privacy applications.\\%
$\bullet$ Cryptographic primitives show promise in enabling practical differentially private applications without a trusted server, while bridging the utility gap between LDP and SDP.
\end{minipage}}

\xh{
\subsection{Improve Accuracy via Anonymous Communication}

In LDP, each data owner independently perturbs their own input (e.g., using the randomized response mechanism) before the aggregation on an untrusted server. This results in a large noise in the final output, $O(\sqrt{N})$ for the case of statistical counting queries~\cite{Cheu2018DistributedDP}. Applications such as Google's RAPPOR~\cite{rappor}, Apple Diagnostics~\cite{appleDP}, and Microsoft Telemetry~\cite{telemetry17} which use this LDP deployment model operate under more practical trust assumptions yet suffer from poor accuracy/utility. Recent works~\cite{prochlo17,shuffleDP18,Cheu2018DistributedDP} show that the use of an anonymous communication channel can help improve the accuracy of statistical counting query for LDP and thereby eliminate the need for a trusted data curator. We will use one of these systems called Prochlo~\cite{prochlo17,shuffleDP18} to illustrate the key idea of how anonymity can help improve the accuracy of such applications.

\subsubsection{Case Study: Prochlo}
Anonymous communication channels, first proposed by Chaum in 1981~\cite{Chaum:1981:UEM:358549.358563}
%\cite{chaum2003untraceable},
are systems that enable a user to remain unidentifiable from a set of other users (called the anonymity set). A larger anonymity set corresponds to a greater privacy guarantee.
Examples of such systems include Mixnets, which use proxies to mix communications from various users.
%Mixnets, which are systems with proxies that mix communications from various users are examples of such systems.
In order to circumvent the limitations of LDP, Google explored the use of an anonymous communication channel to improve the accuracy of queries under differential privacy. The proposed technique is called \emph{Prochlo}~\cite{prochlo17,shuffleDP18}. This technique consists of three steps as shown in Fig.~\ref{fig:crypto_for_dp}B: Encode, Shuffle, and Analyze (ESA). The first encoding step is similar to LDP where data owners randomize their input data independently. The second step uses an anonymous communication channel to collect encoded data into batches during a lengthy time interval and shuffles this data to remove the linkability between the output of the communication channel and the data owners. Last, the anonymous, shuffled data is analyzed by a data analyst.

The shuffling step is the crucial link in achieving anonymous communication by breaking linkability between the user and their data. This step strips user-specific metadata such as time stamps or source IP addresses, and batches a large number of reports before forwarding them to data analysts. Additional thresholding in this step will discard highly unique reports (e.g. a long API bit-vector) to prevent attackers with sufficient background information from linking a report with its data owner. Hence, attacks based on traffic analysis and longitudinal analysis can be prevented, even if a user contributes to multiple reports. Prochlo implements this shuffling step using trusted hardware \ch{as proxies} to eliminate the need for a trusted third party. Furthermore, this shuffling step can amplify the privacy guarantee of LDP and hence improves the accuracy of the analysis, even when there is a single invocation from a user. We will next show the intuition for this base case.

\subsubsection{Accuracy Improvement}
To illustrate how anonymous communication can help improve accuracy, let us look at a simple example of computing the sum of boolean values from $N$ data owners, $f:\sum_{i=1}^{N} x_i$, where $x_i\in \{0,1\}$. In LDP, each data owner reports a random bit with probability $p$ or reports the true bit with probability $1-p$ to achieve $\epsilon$-LDP. When using additional anonymous communication channels, the data owners can enhance their privacy by hiding in a larger set of $N$ values, since the attackers (aggregator and analyst) see only the anonymized set of reports $\{\tilde{x}_1,\ldots,\tilde{x}_N \}$.
%, whose distribution can be simulated using only the aggregate result $f(\tilde{x}_1,\ldots,\tilde{x}_N)$.
The improved privacy guarantee can be shown equivalent to a simulated algorithm that (a) first samples a value $s$ from a binomial distribution $B(N,p)$ to simulate the number of data owners who report a random bit, and then (b) samples a subset of responses for these $s$ data owners from $\{\tilde{x}_1,\ldots,\tilde{x}_N\}$.
 %The parameter $s$ simulates the number of data owners who report a random bit and the subset $H_s$ represents the set of corresponding data owners.
The randomness of these sampling processes can amplify the privacy parameter based on a well studied sub-sampling argument~\cite{Kasiviswanathan:2011:WLP:2078965.2078976,NIPS2018_7865}. Therefore, given the value of the privacy parameter, the required noise parameter %$p$
can be scaled down and hence the corresponding error can be reduced to $O(\sqrt{\log(N)})$. For general bounded real-valued linear statistics, the error is established to be $O(\log(N))$~\cite{shuffleDP18,Cheu2018DistributedDP}. Note that these accuracy improvements assume that there is no collusion between the analyst and the anonymous communication, otherwise, the privacy guarantee will fall back to the same as LDP.

In reference to Fig.~\ref{fig:crypto_for_dp}, these works demonstrate the improvement in going from Fig.~\ref{fig:crypto_for_dp}A to Fig.~\ref{fig:crypto_for_dp}B showing a trade-off between accuracy and trust assumptions.}

\xh{
\subsection{Improve Trust via Encryption \& Secure Computation}\label{sec:en_mpc}

SDP requires the use of a trusted data aggregator to achieve high accuracy. A number of works have explored the use of encryption and secure computation to eliminate the need for this trusted data aggregator~\cite{encryptedDB18, djoin12, SMCQL17}. The key challenge here is to maintain the same level of accuracy as in SDP. We will use one of these proposed systems called DJoin to demonstrate the use of secure computation to enable high accuracy computation without the need for a trusted data aggregator. \ch{There is a complementary synergy between secure computation and differential privacy and thus their combination achieves a strong privacy protection.
%which allows combining them together to achieve strong privacy protection.
%For instance, secure computation reveals the outputs of the computation to all the parties and differential privacy can protect privacy of individual entities involved in the computation while releasing the computation output, %can also be used to reveal these outputs in a privacy conscious manner,
For instance, secure computation ensures all parties learn only the output of the computation but nothing else while differential privacy bounds the information leakage of individuals in the output of the computation, resulting in a system that is better than the use of secure computation or DP alone.
}

\subsubsection{Case Study: DJoin}
Consider a simple setting where two parties would like to compute the intersection size of their data while preserving differential privacy for both datasets. If each party does not trust each other, how can we ensure a constant additive error as if they trust each other? It is well known that the lower bound for this query is $\sqrt{N}$, where $N$ is the data size of each party~\cite{mcgregor2010limits}, if we want to ensure the view of each party satisfies differential privacy. However, if we assume both parties are computationally bounded, a constant additive error can be achieved.

DJoin~\cite{djoin12} offers a concrete protocol for achieving DP under this assumption. This protocol applies private set-intersection cardinality technique to privately compute the noisy intersection set of the two datasets. First, party A defines a polynomial over a finite field whose roots are the elements owned by A. Party A then sends the homomorphic encryptions of the coefficients to party B, along with its public key. Then the encrypted polynomial is evaluated at each of Party B's inputs, followed by a multiplication with a fresh random number. The number of zeros in the results is the true intersection size between A and B. To provide differential privacy, party B adds a number of zeros (differentially-private noise of $O(1)$ independent of data size) to the results and sends the randomly permuted results back to party A. Party A decrypts the results and counts the number of zeros. Party A also adds another copy of differentially private noise to the count and sends the result it back to party B. In other words, both parties add noise to their inputs to achieve privacy. However, the final protocol output has only an error of $O(1)$, which is the same as the SDP setting.

\subsubsection{Trust Improvement}
Using secure computation and encryptions achieves a constant additive error like SDP and prevents any party from seeing the other party's input in the clear. However, this requires an additional assumption of all parties being computationally bounded in the \chh{protocol. Hence, }the type of differential privacy guarantee achieved in DJoin is known as computational differential privacy~\cite{Mironov:2009:CDP:1615970.1615981}. In addition, most of the existing protocols consider honest-but-curious adversaries who follow the protocol specification or consider malicious adversaries with an additional overhead to enforce honest behaviour i.e., verify that the computation was performed correctly.

In reference to Fig.~\ref{fig:crypto_for_dp}, these works demonstrate the improvement in going from Fig.~\ref{fig:crypto_for_dp}D to Fig.~\ref{fig:crypto_for_dp}C eliminating the need for a trusted data aggregator.}
%More complicated model (two parties non-coluding), extend to multi-party. Compare to CDP, the improvement and discussion on the malicious setting and people's work on that.

%In all the protocols above, parties involved in the secure computations are assumed honest but curious and computationally bounded. Given this computational assumption, the privacy guarantee achieved by these protocols is known as \emph{computational} differential privacy.
}

\section{Differential Privacy for Cryptography}\label{sec:dpforcrypto}

\begin{figure*}[!h]
\centering
\includegraphics[width=0.95\linewidth]{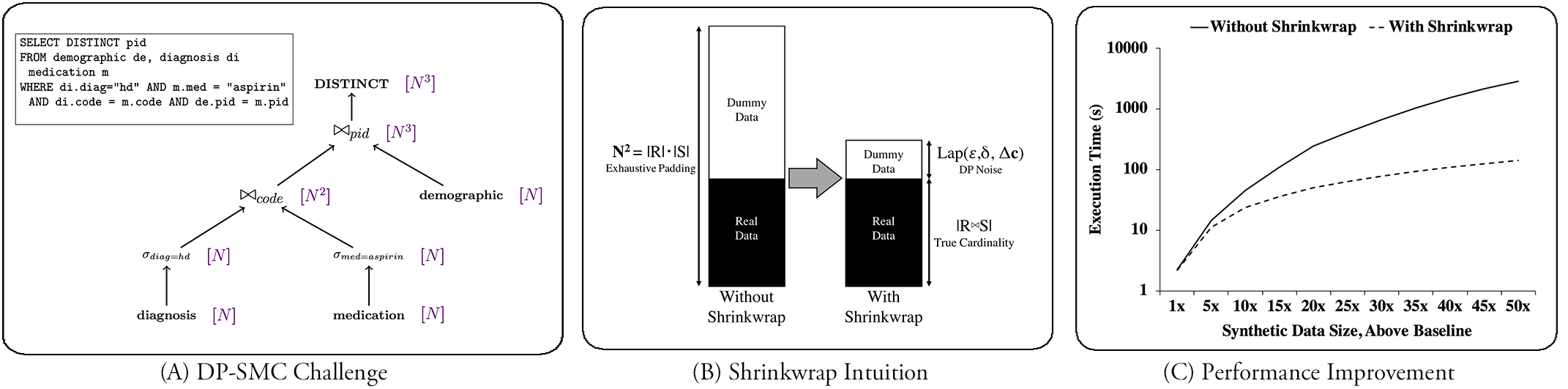}
\caption{(A) Exhaustive padding of intermediate results in an oblivious query evaluation; (B) Effect of Shrinkwrap on intermediate result sizes when joining tables R and S; (C) Aspirin count with synthetic data scaling. Executed using Circuit model. $\epsilon$ = 0.5, $\delta$= .00005.}
\label{fig:shrinkwrap}
\end{figure*}

We have seen in Section~\ref{sec:cryptofordp} that cryptographic primitives show promise in bridging the utility gap between SDP and LDP. However, the large overhead of implementing these conventional cryptographic primitives forms a bottleneck for the deployment of such systems. This motivates the need to enhance the performance of such cryptographic primitives. %Being fundamental privacy enhancing technologies themselves, any improvement in the performance of such primitives directly impacts the practicality of larger systems build on top of these primitives. 
We ask the following question:
\begin{displayquote}
Can we formulate leaky versions of cryptographic primitives for enhancing system performance while rigorously quantifying the privacy loss using differential privacy? 
\end{displayquote}
\ch{\termic primitives are significant for two reasons. First, since the final privacy guarantees of such systems are differential privacy, it is natural to relax the building blocks such as cryptographic primitives to provide differentially private guarantees. Secondly, the composability properties of differential privacy allow for rigorous quantification of the privacy of the end-to-end \chh{system. We} showcase benefits of ``\termic'' systems through two detailed case studies
%on the following three works:}
on (a) secure computation % (2) secure access, 
and (b) secure communication.}

\ch{
\textbf{Key Challenges: }Cryptographic primitives provide strong privacy guarantees. However, deployment of certain cryptographic primitives in practical systems is limited due to the large overhead of these primitives. Relaxing the privacy guarantees in a manner that is amenable to rigorous quantification is difficult and differential privacy can be well utilized to provide a solution to this problem to improve performance overhead.}\\

\hspace*{-\parindent}%
\fbox{\begin{minipage}{0.95\columnwidth}
\textbf{Key Insights of using DP for Cryptography}\vspace{1mm} \\
$\bullet$ We can obtain practical cryptographic implementations that are efficient, while bounding the privacy leakage using differential privacy.\\
$\bullet$ In the context of the end goal of differentially private systems, it is natural to relax the privacy of cryptographic primitives to provide differentially private guarantees.
\end{minipage}}

\subsection{Improve Performance of Cryptographic Computation Primitives}
Cryptographic computation primitives such as Fully Homomorphic Encryption (FHE) and secure Multi-Party Computation (MPC) enable private computation over data. Over the past few years, there has been tremendous progress in making these primitives practical -- most promising of which has been Multi-Party Computation. MPC allows a group of data owners to jointly compute a function while keeping their inputs secret.
%In this section we look at MPC based private computation and in particular, focus on the case study of private query processing.
\ch{
In this section, we show the performance improvement on MPC based private computation, in particular, \emph{differentially private query processing}.}

\comment{
\subsubsection{Key Challenges}
MPC based solutions for private computation show promise in obtaining accurate query answers. The key challenge is their high performance cost in terms of computational and communication overhead. For example, state-of-the-art systems that use secure computation to answer queries with formal privacy guarantees have been shown to bring 4-6 orders of magnitude slowdown compared to non-private systems~\cite{SMCQL17,Xu:2018:UDP:3193111.3193115}. One of the key factors for this slowdown is that the necessity of worst-case running time for avoiding  information leakage due to early termination. This motivates a differentially private approach to reduce the performance overhead to enable practical deployments.
}

\newcommand{\sysname}{Shrinkwrap\xspace}

\xh{
%\subsubsection{Case Study: Differentially Private Query Processing}
\subsubsection{Case Study: Shrinkwrap}
%\todo{Edit this to be more accessible to a wider audience?}

\sysname~\cite{shrinkwrap18} is a system that applies differential privacy throughout an SQL query execution to improve performance. In secure computation, the computation overheads depend on the largest possible data size so that no additional information is leaked. For example, two parties would like to securely compute the answer for the SQL query shown in Figure~\ref{fig:shrinkwrap}A. This query asks for the number of patients with heart disease who have taken a dosage of ``aspirin''. Figure~\ref{fig:shrinkwrap}A expresses this query as a directed acyclic graph of database operators. For example, the first filter operator takes $N$ records from the two parties and outputs an intermediate result which has patients with heart disease (hd). To hide the selectivity (fraction of records selected) of this operator, the baseline system needs to pad the intermediate result to its maximum possible size, which is the same as the input size.  Exhaustive padding will also be applied to the  intermediate output of the two joins and result in an intermediate result cardinality of $N^3$ and a high performance overhead. However, if the selectivity of the filter is $10^{-3}$, cryptographic padding adds a $1000\times$ overhead. Is there a way to pad fewer dummies to the intermediate result while ensuring a provable privacy guarantee?

\sysname helps reduce this overhead by padding each intermediate output of the query plan to a differentially private cardinality rather than to the worst case. As shown in Figure~\ref{fig:shrinkwrap}B, without \sysname, the output of a join operator with two inputs, each of size $N$ is padded to a size of $N^2$. With \sysname, the output is first padded to the worst size and the output is sorted such that all the dummies are at the end of the storage. This entire process is executed obliviously. Then \sysname draws a non-negative integer value with a general Laplace mechanism~\cite{shrinkwrap18} and truncates the storage at the end. This approach reduces the input size of the subsequent operators and thereby their I/O cost. We can see from Figure~\ref{fig:shrinkwrap}C that \sysname provides a significant improvement in performance over the baseline without DP padding for increasing database sizes.

%\subsubsection{Differentially-Private Query Processing}
The relaxed privacy in the secure computation of Shrinkwrap can be quantified rigorously~\cite{shrinkwrap18} using computational differential privacy. Assuming all parties are computationally bounded and work in the semi-honest setting, it can be shown that data owners have a computational differentially private view over the input of other data owners; when noisy answers are returned to the data analyst, the data analyst has a computational differentially private view over the input data of all the data owners.

}

\comment{
\begin{figure*}[t!]
  \centering
  \includegraphics[width = 0.95\textwidth]{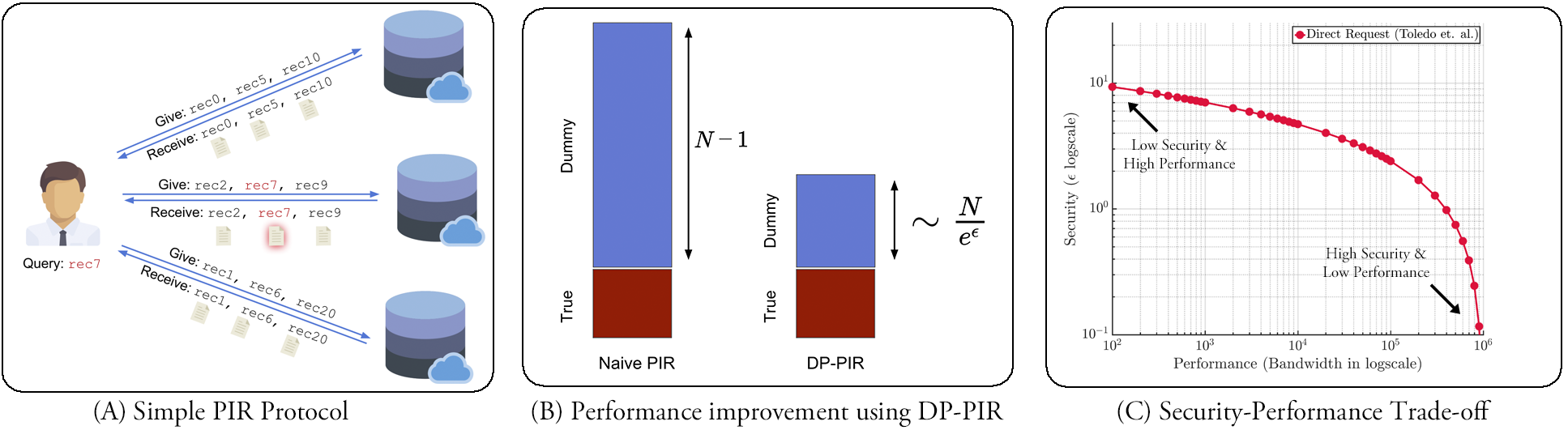}
  \caption{Private Information Retrieval (PIR) is a primitive for private access of data from one or more individually untrusted public servers. PIR can be achieved by hiding the users' true request with a number of artificially generated dummy requests. This results in additional communication overhead equal to the number of dummy requests. Differentially Private PIR protocols aim to reduce this overhead at the cost of differentially private information leakage.}
  \label{fig:dp_pir} 
\end{figure*}

\subsection{Cryptographic Access Pattern Primitives}
Another important security primitive in building secure systems is the hiding of access patterns. Private Information Retrieval (PIR) and Oblivious Random Access Memory (ORAM) are two such cryptographic primitives -- to enable private access to data. They are useful primitives to obtain records from private databases such as a users' Dropbox or  Google Drive (ORAMs) or from public databases such as news websites or DNS servers (PIR). 

\comment{
\subsubsection{Key Challenges?}
Protocols providing access pattern privacy frequently rely on hiding the user's true request among other artificially generated requests (Figure~\ref{fig:dp_pir}A). This introduces an overhead on using such protocols. Despite significant research efforts, state-of-the-art schemes offering perfect access pattern privacy incur large overheads hindering their practical deployment. This invokes a natural question: is it possible to boost the performance of these cryptographic primitives at the cost of statistical privacy? Recent work such as~\cite{DPPIR, wagh2018differentially} answer this question affirmatively.}

\subsubsection{Case Study: Differentially Private Information Retrieval}
We focus on a case study of Differentially Private PIR schemes (DP-PIR)~\cite{DPPIR} to see the promise of relaxed security primitives in practical system deployment. %Differentially private PIR schemes were first formalized by Toledo~\etal~\cite{DPPIR}. 
The set-up consists of multiple non-colluding PIR databases, each holding the same database $D$. The client is interested in a particular item from the database (true request). To make the true request private, the client also generates $p-1$ other ``dummy'' requests (where $p$ is a design parameter) and splits these $p$ requests uniformly across the multiple PIR databases. The PIR databases then respond with the set of requested queries as shown in Fig~\ref{fig:dp_pir}A.

The privacy offered by the above scheme can be quantified using differential privacy. Since the user's true request is masked with $p-1$ other requests and $p \neq N$, the size of the database, the mechanism does not guarantee perfect privacy. At the same time, the overhead of the mechanism defined as the number of database records transferred for each true request is greatly reduced. This reduction in overhead can be seen in Fig.~\ref{fig:dp_pir}B. Similarly, Fig.~\ref{fig:dp_pir}C shows new design space for constructing PIR protocols with various privacy-performance trade-offs. Wagh~\etal~\cite{wagh2018differentially} similarly propose the notion of a differentially private ORAM protocol that achieves faster performance at the cost of relaxed security. A DP-ORAM scheme in conjunction with trusted hardware enables even more efficient DP-PIR schemes.

}

\begin{figure*}[t!]
\centering
\includegraphics[width=17cm]{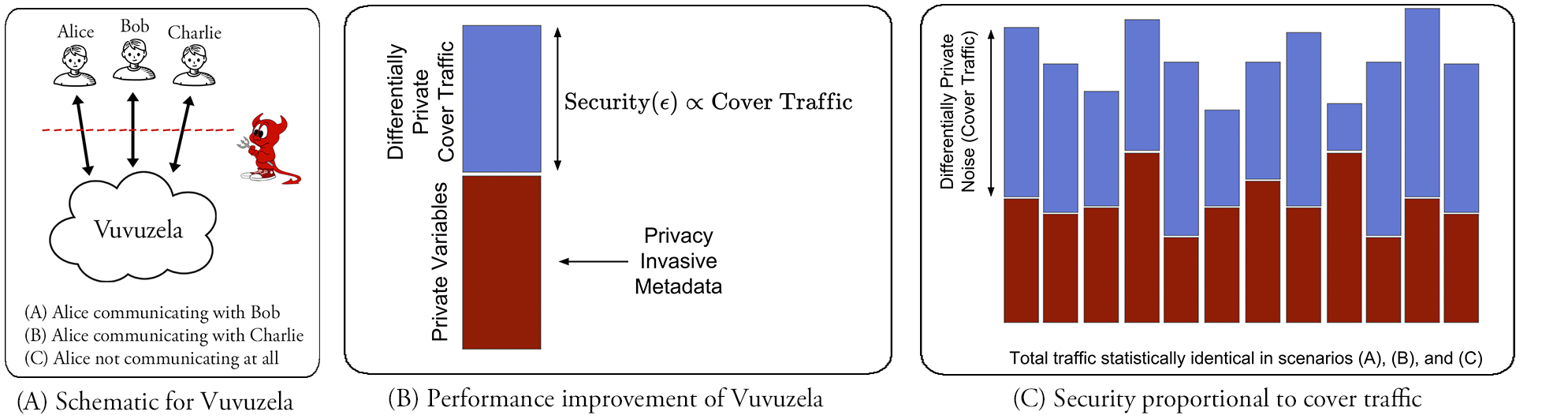}
\caption{Vuvuzela is a secure messaging system. An adversary who can observe and tamper with all network traffic cannot distinguish whether Alice is messaging Bob, Charlie, or is simply not communicating. Vuvuzela uses differential privacy to add noise and mask the privacy invasive metadata, thereby provably hiding information about user communication patters. Vuvuzela achieves a throughput of 68,000 messages per second for a million users scaling linearly with number of users.}
\label{fig:vuvu}
\end{figure*}

\subsection{Improve Performance of Cryptographic Communication Primitives}

%Anonymous communication has been pioneered by the work of Chaum with the introduction of mix-nets and DC-nets. 
Anonymous communication systems aim to protect user identity from the communication recipient and third parties.  Despite considerable research efforts in the domain, practical anonymous communication over current internet architecture is proving to be a challenge. Even if the message contents are encrypted, the packet metadata is difficult to hide. On one end, systems such as Dissent~\cite{dissent} offer strong privacy guarantees yet can scale only to a limited number of participants. On the other end, practical deployed systems such as Tor are vulnerable to traffic analysis and other attacks, limiting their use due to the non-rigorous nature of their privacy guarantees.
\ch{We will show a case study that uses differential privacy to reduce the communication cost while offering rigorous privacy guarantee. We denote this primitive \emph{differentially private anonymous communication.}}

\comment{
\subsubsection{Key Challenges?}
Achieving communication privacy even with a single trusted server/proxy is non-trivial - an adversary observing network traffic can easily infer the communicating parties via traffic analysis attacks such as~\cite{shmatikov2006timing}. It is important that the communication patterns of different users are indistinguishable to an adversary. Achieving this in a system which scales well with the number of users is difficult. The key question yet again is: can differentially private anonymous communication primitives yield significant performance improvements to enable practical systems? }

\ch{
%\subsubsection{Case Study: Differentially-Private Anonymous Communication}
\subsubsection{Case Study: Vuvuzela}
}
Vuvuzela~\cite{vuvuzela} is an anonymous communication system that uses differential privacy to enable a highly scalable system with relaxed yet rigorously quantified privacy guarantees. Vuvuzela provides indistinguishable traffic patterns to clients who: (a) are actively communicating with other clients (b) are not communicating with anyone. In reference to Fig.~\ref{fig:vuvu}, an adversary is unable to distinguish the following three scenarios (a) Alice not communicating (b) Alice communicating with Bob (c) Alice communicating with Charlie. In each of the scenarios, a \chh{Vuvuzela client's network traffic} appears indistinguishable from the other scenarios.

\ch{Vuvuzela employs a number of servers $S_1, \hdots S_n$ where at least one of the servers is assumed to be honest. Clients send (and receive) messages to (and from) the first server, which in turn is connected to the second server and so on. The client creates a layered encryption of its message $m$ i.e., $\mathsf{Enc}_{S_1}(\hdots \mathsf{Enc}_{S_n} (m))$, where $\mathsf{Enc}_{S}(\cdot)$ is the encryption under the key of server $S$. The clients leave messages at virtual locations in a large space of final destinations (called dead drops), where the other legitimate client can receive it. To hide if a client is communicating or not, a client not in an active conversation makes fake requests to appear indistinguishable from a client in an active conversation. If two clients are in active conversation, they exchange messages via the same random dead drop. %whereas each client doing a fake request posts their message to a random dead drop.

Vuvuzela's threat model assumes that at least one server is honest and the adversary is a powerful network level adversary (observing all network traffic) potentially corrupting all other servers\footnote{\ch{Even Tor, a practical anonymous communication system, does not protect against such network level adversaries~\cite{yixinraptor}.}}. The only computation hidden from the adversary is the local computation performed by the honest server \chh{which unlinks users' identifiers} from the dead drops and adds cover (dummy) traffic. As a consequence, the adversary can only observe the number of single or double exchange requests at the dead drop locations. Each Vuvuzela server adds cover traffic using a Laplace distribution to randomize the (a) number of single dead drops and (b) number of double dead drops, which is observable by the adversary. Such random cover traffic addition along with the assumption of at least one honest server provides differentially private guarantees for the observed variables. In other words, Vuvuzela adds noise (cover network traffic) to the two observables (by the adversary) viz. the number of dead drops with one exchange request, and the number of dead drops with two exchange requests, thereby providing communication privacy to clients. This privacy relaxation enables Vuvuzela to scale to a large number of users -- it can achieve a throughput of 68,000 messages per second for a million users. Systems such as Stadium~\cite{stadium}, and Karaoke~\cite{lazar2018karaoke} further improve upon Vuvuzela and scale to even larger sets of users.} %In fact, all systems leak some information, quantified by differential privacy but showcase improvement in performance. }

%
%
%then uses differential privacy to add cover traffic such that 
%
%
%Vuvuzela hides the public observables of the adversary by adding dummy traffic to the network. }
%
%Depending on whether the client is not communicating or communicating with another client, the traffic connections result in single or double exchange requests respectively at dead drop locations. This is publicly observable by an adversary. Vuvuzela adds noise (dummy network traffic) to hide these two observables viz, the number of dead drops with one exchange request, and the number of dead drops with two exchange requests, and thereby provides communication privacy to clients. This privacy relaxation enables Vuvuzela to scale to a large number of users -- it can achieve a throughput of 68,000 messages per second for a million users. Systems such as Stadium~\cite{stadium}, and Karaoke~\cite{lazar2018karaoke} further improve upon Vuvuzela and scale to even larger sets of users. In fact, all systems leak some information, quantified by differential privacy but showcase significant gains in performance. 

\subsection{Limitations of Differentially Private Cryptography}
To end our discussions, we caution readers against \chh{careless} combinations of differential privacy and cryptographic primitives. First, the limitations of both differential privacy as well as cryptographic primitives apply to \termic primitives. For instance, an open question is deciding an appropriate level for the privacy budget. \ch{Most applications that utilize DP to improve the performance of cryptographic systems involve a trade-off between the level of privacy achieved and the performance of the systems. More generally, differentially private cryptographic systems open up new trade-offs in a privacy-performance-utility space. 
%For instance, in the case of DP-PIR, using a high privacy value directly leads to higher performance overhead for the system (privacy-performance trade-off). However, perfect utility is still maintained, i.e., DP-PIR is also a PIR protocol.
For instance, in the case of Shrinkwrap, weaker privacy guarantee directly leads to lower performance overhead (privacy-performance trade-off while keeping the accuracy level of the query answer constant). On the other hand, systems such as RAPPOR allow for approximate computation of statistics and primarily provide a privacy-utility trade-off.} Second, designers need to carefully consider the suitability of these hybrid techniques in their applications as these combinations involve more complex trust assumptions and hence a more complicated security analysis. \chh{We remind the reader that while proposing newer differentially private systems for cryptography, it is imperative to understand the meaning of the privacy guarantees for the application in context. In other words, differentially privacy for cryptography may not be the right thing to do in all cases; however, it is well motivated when the goal is to build a differentially private system.} Finally, composition results, which bound the privacy loss for a sequence of operations need to be independently studied. 

%We would like to remind the reader that while proposing newer differentially private systems for cryptography, it is imperative to understand the meaning of the privacy guarantees for the application in context. In other words, DP for Cryptography may not be the right tool in all cases; however, it is well motivated when the goal is to build a differentially private system.

%

%\subsection{DP-Cryptographic Primitives}
%The above works strongly hint at the promise of using differential privacy to provide practical solutions to problems in applied cryptography. In particular, perfectly secure cryptographic primitives such as secure computation, PIR, ORAM, and anonymous communication have limitations in deployment given the scale and complexity of real world data. The use of relaxed privacy notions of such primitives, seen through the examples of ShrinkWrap, DP-PIR and Vuvuzela, have shown that significant performance improvements can be achieved at the cost of relaxed privacy quantified using differential privacy. The composition properties of differential privacy enable a rigorous privacy analysis in complex systems building on many such \termic primitives. 
%
%Section~\ref{sec:cryptofordp} shows how cryptographic primitives can improve the accuracy of differentially private queries in complex deployment models such as LDP. The composability properties of differential privacy imply that efficient \termic primitives can be used synergistically in complex deployment scenarios of differential privacy.
%

\section{Discussion and Open Questions}\label{sec:open}

In this section, we provide directions for future work highlighting important and emerging open questions in the field. We discuss open challenges in deploying differential privacy in the real world -- realistic datasets, alternative models and trust assumptions, and other \termic primitives. Finally, we caution readers against callous combinations of differential privacy and cryptography.

%\subsection{Research Directions: Differentially Private Cryptographic Primitives}
%In Sections~\ref{subsubsec:SDP} and~\ref{subsubsec:LDP}, we have compared two popular models of deployment of differential privacy. Simply stated, there is a gap in the trust assumptions and the utility achieved by mechanisms in SDP and LDP - high trust assumptions, high utility in SDP and lower trust assumptions, lower utility in LDP. An important emerging direction of research has been to explore the role of cryptography to bridge the trust-accuracy gap (cryptography for DP) and obtain the best of both worlds: high accuracy without assuming trusted data aggregator. For instance, Fig.~\ref{fig:crypto_for_dp}B shows how a Mixnet can be used to bridge the utility gap in differential privacy - from a $O(\sqrt{N})$ noise addition requirement to a $O(\log N)$ noise addition. Another important research direction has been to relax the cryptographic guarantees motivated by the context of DP applications, and instead enhance the performance of these primitives (DP for cryptography). For instance, Fig~\ref{fig:shrinkwrap}C shows the improvement in execution time of an oblivious query evaluation over distributed datasets by relaxing privacy. Next, we explore these two threads of research -- using cryptography for differential privacy and using differential privacy for cryptography -- highlighting key results in each. 

\textbf{Differential Privacy Frameworks -- SDP, LDP, and Beyond:} 
Over the past decade, there has been significant progress in enabling applications in the standard differential privacy model. For instance, there have been research efforts in attuning differential privacy to handle realistic challenges such as multi-dimensional and complex data -- involving graphs, time series, correlated data~\cite{changchangDDP, kasiviswanathan2013analyzing}. Similarly, there has been work in designing a tailored differential privacy mechanism that is optimized for particular application setting to achieve good accuracy~\cite{HDMM, PrivateSQL}. Prior work has explored combinations of sequential and parallel composition, dimensionality reduction, and sensitivity bound approximations to achieve good accuracy in the SDP model. \ch{However, much work needs to be done in adapting state-of-the-art techniques in SDP to more complex deployment scenarios such as LDP. For instance, an open question is the following:
\begin{displayquote}
Is there an algorithm that can efficiently search the space of differentially private algorithms in the LDP setting for the one that answers the input query with the best accuracy?
\end{displayquote}
Research advances have demonstrated such mechanisms for the SDP model~\cite{HDMM, PrivateSQL}, however, the discovery of such mechanisms in the LDP setting remains an open question.
%\ch{For instance, a concrete open question to the community is to come up with an automated tool to apply an optimized differential privacy mechanism in the LDP setting. This has been done for SDP~\cite{HDMM, PrivateSQL}. A concrete open question is the following:
%\begin{displayquote}
%Is there an algorithm that can efficiently search the space of differentially private algorithms for one that answers the input query with the best accuracy?
%\end{displayquote}
%it is an important open question to generate automated optimized differential privacy mechanisms for answering queries in the LDP setting. 
On a similar note, it is unclear how nuanced variants of differential privacy that have been proposed to handle these more complex databases~\cite{changchangDDP, kasiviswanathan2013analyzing} in the SDP setting translate into LDP or more complex deployment settings.}

\textbf{Differential Privacy in Practice -- Trust Assumptions vs Accuracy Gap:}
We have seen how deployments of differential privacy that differ in the trust assumptions provide roughly the same privacy guarantee, but with varying levels of accuracy. In particular, we looked at a two popular deployment scenarios viz., SDP and LDP. There exist other trust assumptions that we have not covered in this article in detail. For instance, Google's recently proposed Prochlo system~\cite{prochlo17} uses trusted hardware assumptions to optimize utility of data analytics. On a similar note, Groce~\etal~\cite{maliciousDP} consider yet another model -- where the users participating are malicious. This is the first work to explore a malicious adversarial model in the context of differential privacy and the development of better accuracy mechanisms for such a model is an open research question. \ch{More concretely, we can ask: 
\begin{displayquote}
What other models of deployment of differential privacy exist and how do we design mechanisms for them? Can other technologies such as MPC, FHE, trusted hardware open up new opportunities in mechanism design?
\end{displayquote}
An interesting theoretical question is to characterize the separation between different trust models in terms of the best accuracy achievable by a differential privacy algorithm under that model. For instance, McGregor~\etal~\cite{mcgregor2010limits} provide separation theorems i.e., gaps in achievable accuracy between (information-theoretic) differential privacy and computational differential privacy for two-party protocols. In reference to Section~\ref{sec:en_mpc} we can ask the following concrete question:
\begin{displayquote}
In the Mixnets model (Fig.~\ref{fig:crypto_for_dp}B), what is the lower bound on the errror for aggregate queries over relational transformations (like joins and groupby) over the data records? An example of such an aggregate is the degree distribution of a graph that reports the number of nodes with a certain degree.
%Is an accuracy of $O(1)$ achieveable for statistical queries in the deployment model using Mixnets? What are the lower bounds for non-statistical queries such as disjunctive normal form (DNF) queries in this model?
%in a model given in Fig.~\ref{fig:crypto_for_dp}C, where the parties are computationally unbounded? If not, can lower bounds be established for what is achieveable accuracy in that deployment model?  \todo{Double check!}

%Is an accuracy of $O(1)$ achieveable in a model given in Fig.~\ref{fig:crypto_for_dp}C, where the parties are computationally unbounded? If not, can lower bounds be established for what is achieveable accuracy in that deployment model?  \todo{Double check!}
\end{displayquote}}

\textbf{Relaxing Cryptographic Security via Differential Privacy:} 
\ch{The emerging paradigm of leaky yet differentially-private cryptography leads to a number of open questions for the research community. \chh{So far, the research community has explored the intersection of differential privacy and cryptographic primitives in limited contexts such as ORAM, MPC, and anonymous communication. However, there exists a broader opportunity to explore the trade-offs of \termic{} primitives in contexts such as program obfuscation, zero-knowledge proofs, encrypted databases, and even traffic/protocol morphing.} As described in Section~\ref{sec:dpforcrypto}, we can ask: 
\begin{displayquote}
\chh{What other cryptographic primitives can benefit in performance from a privacy relaxation quantified rigorously using differential privacy? How can we design such relaxed primitives?}
%\chh{What other cryptographic primitives can benefit in performance from a privacy relaxation quantified rigorously using differential privacy?}
%Can we design general mechanisms to improve the performance of cryptographic primitives by relaxing the privacy guarantees to differential privacy? 
\end{displayquote}
In the context of differentially-private data analysis, there is a trade-off between privacy and utility. In the context of differentially-private cryptographic primitives and resulting applications, there is a broader trade-off space between privacy, utility, and performance. Another open question is the following:
\begin{displayquote}
What lower bounds exist for overhead of cryptographic primitives when the privacy guarantees are relaxed using differential privacy?
%rethink the lower bounds for overhead of cryptographic primitives when cryptographic guarantees are relaxed using differential privacy
\end{displayquote}
Another challenge is how to design optimized protocols that achieve desired trade-offs in the new design space of differentially-private cryptography. \chh{The trade-off space between privacy, utility, and performance is non-trivial, especially for complex systems. An interesting research question is:
\begin{displayquote}
How to correctly model the trade-off space of real systems so that system designers can decide whether it is worth sacrificing some privacy or utility for a better performance?  
\end{displayquote}}}

%For example, conventional PIR schemes have perfect privacy but poor performance (i.e., poor utility). However, differentially-private PIR provides a new design point, weaker privacy yet better performance. Furthermore, recent work has shown that a DP-PIR protocol can be combined with a perfect anonymity system to further improve this privacy-utility trade-off achieved by the DP-PIR mechanism~\cite{DPPIR}.
%This also raises another interesting research question:
%\begin{displayquote}
%What are achievable limits in the privacy-utility-performance space opened by differentially private mechanisms for a given technical system?
%\end{displayquote}}

\comment{
\section{Conclusions}\label{sec:conclusions}
In this work, we have synthesized research at the intersection of differential privacy and applied cryptography. We categorize work in this domain into two main categories (1) cryptography for differentially private applications and (2) differentially privacy for cryptographic primitives. The former explores the use of cryptographic primitives to enable differentially private applications in models with more practical trust assumptions such as LDP. The latter deals with relaxations of cryptographic primitives that enhance performance at the cost of reduced yet differentially private information leakage. This unique perspective opens a number of important questions which can enable the next generation of privacy-preserving systems.
}

\section{Acknowledgments}
This work was supported by the National Science Foundation under grants 1253327, 1408982, CIF-1617286, and CNS-1553437; the Army Research Office YIP award; and by DARPA and SPAWAR under contract N66001-15-C-4067.

\bibliographystyle{abbrv} % standard abbrv style
\bibliography{main}  % substitute the name of 'your' Bibliography here
\balancecolumns
\end{document}